# Teaching vs. Learning: Changing Perspectives on Problem Solving in Physics Instruction

William J. Gerace & Ian D. Beatty

*Scientific Reasoning Research Institute & Department of Physics*
*University of Massachusetts Amherst*
*Amherst, Massachusetts, USA*

Problem solving is central to physics instruction. Results from Physics Education Research (PER), however, demonstrate that traditional ways of teaching with problem solving are inefficient and ineffective for promoting true physics expertise. PER findings give rise to a perspective on physics expertise, learning, and problem solving that can illuminate the reasons why problem solving in traditional instruction fares poorly and suggest remedies. At the heart of the remedies lies a rethinking of the instructional model in which teachers focus less on presenting subject material and more on engineering learning experiences and guiding students' learning efforts, while students strive to become active, self-monitoring constructors of knowledge.

## Introduction

Problem solving plays a central role in traditional physics instruction. Teachers and textbooks work example problems to illustrate concepts and principles, to demonstrate procedures, and to clarify points of likely confusion. Course work includes homework problems to direct students' studying, help them to figure out ideas, and provide skills practice. Course exams consist of problems to assess student learning. And high-stakes exams on standardized tests and qualifying exams vet applicants to postgraduate degree programs and determine whether they may proceed.

Physics educators use problem solving to promote learning and to measure mastery of physics. But is it really effective for these? A wealth of evidence from Physics Education Research (PER) suggests that problem solving, as commonly employed, is far less effective than many hope or believe. PER results also indicate how instruction, including the use of problem solving, can be reconceived and reshaped to strengthen physics teaching at the high school and university levels.

# The Effectiveness of Problem Solving in Traditional Physics Instruction

Before entertaining the question of whether problem solving as traditionally employed in physics instruction is effective, we must ask: effective at what? What are the ultimate goals that we hope to achieve with our instruction?

We think most educators would agree that we aspire to more than keeping our students busy and then having them pass our final exams. Rather, we desire our students to progress towards *expertise* in physics. We want them to develop the ability to apply physics concepts to physical situations and reason with them, and to solve a broad range of problem types including novel ones not practiced during class. At the end of the course, we would like our students to understand (not just recall) the material we've covered, and—to use a term from current educational theory—to *transfer* the ideas and skills they've gained to new contexts.

The field of Physics Education Research has seen considerable growth over the last thirty years or so. Its practitioners have amassed a convincing array of evidence that traditional physics instruction, emphasizing derivations and problem solving, is rather inefficient at achieving the above objectives. The approach does help some students learn most of what we'd like as well as we'd like, and helps many students learn some of what we'd like to some degree, but in general it serves our students poorly and is not well matched to our avowed goals. This is not the venue for a thorough literature review, but a brief summary of the nature and scope of the evidence is in order.

The most statistically compelling evidence comes from the use of instruments such as the *Force Concept Inventory* (FCI), which evaluates understanding of basic concepts in introductory mechanics and probes for the presence of common misconceptions (Hestenes *et al.*, 1992). It has been constructed to reveal whether students "get" ideas such as Newton's Third Law at an intuitive level, not whether they can solve formal and quantitative problems. The FCI was carefully crafted after extensive interview-based research into students' learning of mechanics ideas, and then thoroughly validated. The instrument is typically given to students as a pre-test and post-test surrounding a physics course to evaluate the improvement in their conceptual understanding.

A study by Richard Hake of FCI data from 62 courses with 6542 students has yielded convincing evidence that traditional lecture-and-homework instruction produces a relatively small average improvement. On the other hand, instruction employing "interactive engagement" methods based on PER and other learning research typically produces an average improvement twice as large (Hake, 1998). Other research with the FCI and similar instruments supports the conclusion that traditional physics instruction is not effective at impacting students' conceptual understanding of the physical world (L. McDermott, 1991, 1993; Mestre, 1991; Redish & Steinberg, 1999). Many instructors are greatly surprised by how poorly their students fare (Mazur, 1977).

These results are supported by evidence from many other avenues of research: detailed studies of the persistence of students' misconceptions and naïve intuitions (Halloun & Hestenes, 1985a, 1985b; Pfundt & Duit, 1994); investigations of students' inability to explain their problem solutions or to construct a written strategy for solving a problem (Clement, 1981; Leonard *et al.*, 1996; Touger *et al.*, 1995); studies of students' ability to perceive the "deep structure" of a problem and to categorize problems by similarity (Chi *et al.*, 1981; P. H. Hardiman *et al.*, 1989a); research into the narrow context dependence of students' knowledge (Mestre, 2002, 2005); and exploration of students' long-term retention of material learned in class (Leonard et al., 1996). Overall, the evidence is so compelling that nobody involved in PER doubts physics instruction is broken and needs fixing.

This raises the question of *why* problem solving, as traditionally employed, does not achieve the aims for which it is intended. Answering this question requires a brief detour to present a perspective on physics expertise, learning, and problem solving skill grounded in cognitive and educational research.

## A Research-Based Perspective on Physics Expertise, Learning, and Problem Solving

### *Expertise*

We wish our students to develop expertise in physics, but what exactly is "expertise"? Its manifestations are clear: expertise in physics permits a person to reason and analyze physical situations in terms of concepts, supplementing such reasoning with quantitative calculations as necessary. It enables one to transfer knowledge and skills to new contexts as they are encountered. More than just knowing how to solve a set of standard problems, an expert has the skills to adapt and attack new and unfamiliar problems (Bransford *et al.*, 1999).

What does an expert learn that underlies such ability? Research has illuminated the ways in which an expert's knowledge differs from a novice's (Chi et al., 1981; Gerace, 1992; Zajchowski & Martin, 1993). One significant area of difference is in the "structure" of the knowledge store: experts don't just have more knowledge, they have organized their knowledge to be useful. Experts have a large store of domain-specific knowledge, whereas novices have a sparse knowledge set with gaps. Experts' knowledge is richly interconnected and interrelated, while novices' knowledge forms disconnected groupings around distinct topics. Experts structure their knowledge hierarchically, organized by fundamental principles, but novices store theirs chronologically as it is learned. Experts integrate multiple representations of ideas, whereas novices often have only one representation, or are unable to relate different representations. As a result, experts have good recall of their knowledge, and can access whatever part is relevant for a problem, while novices have poor recall and don't have access to a particular bit unless cued by something (perhaps a familiar, standard problem type) they have been drilled to associate it with.

Experts and novices also differ in their problem solving behaviors. Experts employ forward-looking concept-based strategies, whereas novices typically employ backward-looking means-ends techniques. Experts often apply qualitative analysis techniques, especially when stuck, while novices manipulate equations. Experts have a variety of tactics for getting unstuck, but novices cannot generally get unstuck without outside help. Experts can think about and monitor their problem solving while engaged in it, but for novices, problem solving consumes all available mental resources. Finally, experts can and generally do check their answers via alternative methods, while novices usually have only one way to solve a problem.

### *Learning*

Research on physics expertise helps us understand the target we want students to reach. To understand how to get them there, we must turn to research on learning.

Modern educational theory is based on the epistemology of *constructivism*: in essence, the recognition that knowledge (as opposed to information) cannot be transmitted or observed, but must be constructed as the result of cognitive processes within the human mind (Gerace, 1992; Smock & von Glasersfeld, 1974; von Glasersfeld, 1998). As a basis for learning, constructivism has the following premises:

1. Knowledge is constructed, not transmitted.
2. Prior knowledge impacts learning.
3. Construction of knowledge requires purposeful and effortful activity.
4. Initial understanding of a concept is local and context-limited, not global.

In the context of formal instruction, these premises mean the following:

1. Students arrive in class with an established view of both the physical world and learning, formed from prior experience and learning.
2. Students' existing world view filters all new observations and experience.
3. Students are emotionally attached to their world views.
4. Students must expend much effort to revise and restructure their world views.

This perspective has deep and pervasive implications for how we conceive of and practice instruction. Rather than concerning ourselves with the clearest and most logical way to present our subject material to a classroom of "blank slates," we must design learning experiences in accord with student's varied and idiosyncratic "initial states." This means we must continually assess what they know and how they are interpreting our instruction. Rather than "teaching," we find ourselves engaged in the design and management of beneficial educational experiences, guiding students as they engage in an iterative, difficult, and often frustrating process of sense-making and knowledge-organizing. Often, we must work hard to overcome the misconceptions and naïve understandings students begin with. We must also repeatedly revisit ideas as students extend their understanding into new contexts and build increasingly sophisticated knowledge structures.

Learning physics is more than just coming to understand the concepts of physics, however. It also entails learning how to think like a physicist: developing the *habits of mind* that allow one to make productive use of the knowledge base (R.J. Dufresne *et al.*, 2000). PER has come to see learning as an active process of engaging in directed cognitive activity to construct useful knowledge structures while practicing skills and mental processes.

### *Problem Solving*

Research has also illuminated the process of problem solving. An individual's problem solving ability depends strongly on the organization, not just the extent, of her knowledge store; frequently, a student possesses the requisite knowledge to solve a problem, but does not think of it without prompting or coaching. A typical physics curriculum leaves students poorly prepared to select physics ideas from their entire repertoire as needed, except when cued by stereotyped problems that they recognize (Gerace *et al.*, 2001; P. T. Hardiman *et al.*, 1989b; Zajchowski & Martin, 1993).

Expert problem solvers faced with a challenging quantitative problem (rather than a standard one they recognize and remember the solution to) proceed through four phases of analysis: *conceptual analysis* (orienting, exploring); *strategic analysis* (planning, choosing); *quantitative analysis* (executing, determining, answering); and *meta-analysis* (reflecting, checking, challenging, relating). In typical instruction, only quantitative analysis is explicitly modeled for students, leaving them to develop the other skills on their own (R.J. Dufresne *et al.*, 2000).

Students do not develop problem solving facility all at once. Rather, they progress through five identifiable but overlapping and interdependent stages of cognitive development: exploration of pre-existing notions and introduction of new ideas; honing and linking of concepts; use of concepts to analyze and reason about situations; organization and prioritization of knowledge; and development of general problem-solving strategies (R.J. Dufresne et al., 2000).

## Why Problem Solving in Traditional Physics Instruction is Ineffective

Using the PER-based perspective on physics expertise, learning, and problem solving developed in the previous section, we can understand why problem solving as it is traditionally employed is ineffective for developing true expertise in students.

Problems typically used in traditional physics instruction are for the most part goal-directed, narrow, disconnected, and simplistic. By "goal-directed", we mean that they give students a very specific objective, such as calculating a physical quantity. By "narrow", we mean that they can be solved by the straightforward application of a single principle, definition, or procedure. By "disconnected", we mean that they are closely related to the topics and worked-out examples recently covered in lecture or assigned readings, and do not integrate previously acquired knowledge. By "simplistic", we mean that they ignore most of the complicated, messy physics that is needed to address real-world situations.

When faced with such problems, students tend to engage in a host of undesirable behaviors rather than in cognitive activity that builds and structures knowledge and develops desirable habits of mind. They focus excessively on the goal of determining the answer. They construct an abstract representation of the problem based primarily on superficial features of the situation, with limited use of concepts. They employ means-ends analysis to determine a solution path, and engage in equation manipulation. And they try to use physics that is familiar, rather than that which is new and unfamiliar.

What students don't do, but should, is: analyze situations in terms of concepts; interpret mathematical formalism; employ multiple representations; seek and weigh alternative solutions; formulate a strategy before solving; compare and contrast with more familiar situations; and monitor and reflect upon their own problem-solving.

## Rethinking Instruction and Problem Solving

At the root of the situation is what we call the *two-tiered problem*:

1. What students *know* determines how they engage in problem solving activities.
2. How students engage in problem solving activities determines what they *learn*.

The central issue here is that *the manner in which students learn is itself learned*. To turn this to our advantage, we introduce a *two-tiered solution*:

1. Structure problem solving activities to influence what students pay attention to.
2. Engage in explicit meta-cognitive communication about learning.

The central issue of this solution is that *students must be made aware of their own learning habits*, promoting them to conscious collaborators in the knowledge construction process.

We accomplish the first tier of the solution by designing learning activities—alternatives to traditional problem solving—via a *model-based design paradigm*. That paradigm begins with a *model* of the knowledge elements, structures, and skills necessary for expert-like problem solving, as illuminated by PER. From the model and an understanding of our students' current knowledge and ability, we determine *knowledge structures and skills* to target for development. From these we infer *cognitive processes* that students can engage in to build the structures and develop the skills. These processes guide the design of *instructional activities* we can teach through. Students' active engagement in these activities results in *learning*.

This paradigm can help us design problems for student problem-solving that, although superficially similar to traditional problems, have a drastically different and more beneficial effect. For example, consider the situation of a ladder leaning against a wall, in which there is friction between the ladder and the floor but not between the ladder and wall. A man stands upright on a single rung. To make a traditional problem, we might specify numerical values for the mass of the ladder and man, the length of the ladder, the angle the ladder makes with the floor, and the position of the man on the ladder, and ask students to calculate the friction force on the base of the ladder. On the other hand, to promote beneficial cognitive processes, we could omit the numerical values and instead ask students whether the friction force on the base of the ladder increases, decreases, or stays the same as the man walks up the ladder. This latter version induces students to engage in conceptual, qualitative, and comparative reasoning, and discourages equation manipulation and means-ends analysis.

A more dramatic option is to present students with a problem and then, instead of having them solve it, asking them write out a strategy (no equations) or to select which of several principles would be most useful for solving the problem. This directs their attention to the development of strategic knowledge, helps them structure knowledge according to key principles and utility, and creates the opportunity for a productive, high-level conversation about strategic problem solving skills.

We accomplish the second tier of the two-tiered solution by discussing cognition and learning alongside of physics, by encouraging students to monitor their thinking and be aware of their learning habits, and by designing activities that direct students' attention to meta-cognitive issues. Meta-cognitive skills *can* be taught, but it takes determined, sustained emphasis and consistency of explicit and implicit messages. In essence, we are redefining our role as an instructor to be more of a learning coach or guide and less of an authority on and presenter of physics content.

Physics Education Research has led to more than the development of novel problem types for use in a traditional problem solving setting. Employed wisely *classroom response systems*—technology for posing questions to students in class, and electronically collecting and displaying their answers—can transform a classroom from a largely one-way lecture to a dynamic discussion with engaged students, even with hundreds of students in the room (Beatty, 2004; Robert J. Dufresne *et al.*, 1996; Mazur, 1977; Roschelle *et al.*, in preparation). *Workshop Physics* entirely replaces lecture with guided laboratory-style exploration of physics ideas (Laws, 1997). *Tutorials* in the University of Washington style supplement lecture with collaborative, process-oriented, inquiry-based development of important concepts (L. C. McDermott & Shaffer, 2002, 2003). Other "reforms" abound, too many to list here (L. C. McDermott & Redish, 1999).

An important component of the model based design paradigm is ascertaining students' idiosyncratic and time-varying needs, confusions, skills, and state of understanding. This calls for the use of *formative assessment* techniques: assessment to guide and enhance, rather than evaluate, learning (Black & Wiliam, 1998a, 1998b; Bransford et al., 1999; Sadler, 1989; Stiggins, 2002). Formative assessments are frequent, brief, low-stakes, and designed to reveal thinking and gaps in knowledge rather than measuring ability.

High-stakes *summative assessment* approaches need rethinking as well. "Assessment drives learning" is a common phrase in educational research circles, since students use exams to determine what they should pay attention to (and instructors use standardized tests to determine what they must cover). If an instructor's exams ask students to recognize common problem types, manipulate equations, and calculate quantitative answers, then no reform of the course curriculum, methods, or assignments will convince them that they should be more concerned with conceptual understanding, qualitative and strategic analysis skills, and meta-cognitive awareness of learning (Bransford et al., 1999).

## Conclusions and Final Thoughts

Problem solving as traditionally practiced within physics education is neither particularly efficient nor effective at helping students develop true expertise. Problem solving can and should remain central to instruction, but *how* it is practiced must change, as part of a fundamental rethinking of the instructional process.

Results from Physics Education Research have illuminated the nature of expertise and the cognitive processes required to achieve it. To be consistent with these findings, instruction must be reformulated as a bidirectional communication process, with students as self-aware, active partners. All course components, including problem solving activity, should be designed to promote cognitive processes that build structured knowledge and develop desirable habits of mind, and to guide students through the five stages of cognitive development. Meta-cognitive skills should be explicitly taught, and the learning of physics should be a first-class topic of instruction alongside the content of physics. Finally, assessment must serve rather than dictate learning. Fortunately, a large and growing array of PER-based techniques and curriculum materials are available to help instructors implement such a transformation.

Accomplishing this dramatic recasting of the instructional model is much harder than describing it, of course. Achieving it is a difficult and extended process for any instructor. It helps to think of ourselves as engineers of learning experiences and as learning guides, to remember that no "perfect" lectures, activities, procedures, or curricula are possible, and to enter the classroom as much to learn about students and teaching as to teach about physics. The better we understand our own learning and thinking — of physics and of teaching — the better we can help students develop their own.

*The astute reader may have noticed that this paper's title is ambiguous. Does "changing perspectives" indicate an intent to inform, or to influence? Successful communication requires problem solving by all parties involved.*